\documentclass{article}

\usepackage{microtype}
\usepackage{graphicx}
\usepackage{subfigure}
\usepackage{booktabs} % for professional tables

\usepackage{kotex}
\usepackage{lipsum} 
\usepackage[dvipsnames]{xcolor}
\usepackage{multirow}

\usepackage{hyperref}

\usepackage[accepted]{icml2020}

\icmltitlerunning{CareCall: a Call-Based Active Monitoring Dialog Agent for Managing COVID-19 Pandemic}

\begin{document}

\twocolumn[
\icmltitle{CareCall: a Call-Based Active Monitoring Dialog Agent \\ for Managing COVID-19 Pandemic}

%CareCall: Phone-Call Active Monitoring Dialog Agent \\ for Plattening the Curve on COVID-19

% ClovaCall: Korean Goal-Oriented Dialog Speech Corpus for Automatic Speech Recognition of Contact Centers

\icmlsetsymbol{equal}{*}

\begin{icmlauthorlist}
\icmlauthor{Sang-Woo Lee}{equal,clova}
\icmlauthor{Hyunhoon Jung}{equal,clova}
\icmlauthor{SukHyun Ko}{clova}
\icmlauthor{Sunyoung Kim}{clova}
\icmlauthor{Hyewon Kim}{clova}
\icmlauthor{Kyoungtae Doh}{clova}
\icmlauthor{Hyunjung Park}{clova}
\icmlauthor{Joseph Yeo}{clova}
\icmlauthor{Sang-Houn Ok}{clova}
\icmlauthor{Joonhaeng Lee}{clova}
\icmlauthor{Sungsoon Lim}{clova}
\icmlauthor{Minyoung Jeong}{clova}
\icmlauthor{Seongjae Choi}{clova}
\icmlauthor{SeungTae Hwang}{clova}
\icmlauthor{Eun-Young Park}{seongnam}
\icmlauthor{Gwang-Ja Ma}{seongnam}
\icmlauthor{Seok-Joo Han}{seongnam}
\icmlauthor{Kwang-Seung Cha}{seongnam}
\icmlauthor{Nako Sung}{clova}
\icmlauthor{Jung-Woo Ha}{clova}
\end{icmlauthorlist}

\icmlaffiliation{clova}{NAVER Corp.}
\icmlaffiliation{seongnam}{Seongnam City}
% \icmlaffiliation{goo}{Googol ShallowMind, New London, Michigan, USA}
% \icmlaffiliation{ed}{School of Computation, University of Edenborrow, Edenborrow, United Kingdom}
\icmlcorrespondingauthor{Nako Sung}{nako.sung@navercorp.com}
\icmlcorrespondingauthor{Jung-Woo Ha}{jungwoo.ha@navercorp.com}

\icmlkeywords{Machine Learning, ICML}

\vskip 0.3in
]



\printAffiliationsAndNotice{\icmlEqualContribution} % otherwise use the standard text.

\begin{abstract}
Tracking suspected cases of COVID-19 is crucial to suppressing the spread of COVID-19 pandemic. Active monitoring and proactive inspection are indispensable to mitigate COVID-19 spread, though these require considerable social and economic expense.
To address this issue, we introduce CareCall, a call-based dialog agent which is deployed for active monitoring in Korea and Japan. We describe our system with a case study with statistics to show how the system works.
Finally, we discuss a simple idea which uses CareCall to support proactive inspection.

\end{abstract}

\section{Introduction}
The situation of COVID-19 pandemic has been so serious that there exist more than 7.5M patients and 420k dead people by early June. Under this serious situation, tracking patient spread is crucial to mitigate COVID-19 pandemic. In particular, individual quarantine and large-scale active monitoring are known to be significantly effective in the pandemic mitigation~\cite{world2020critical,weissleder2020covid}. To maintain social quarantine and gather symptoms in real-time, however, enormous social and economic expenses are required due to the considerable cost of testing performed by a restricted number of medical staff with limited infection testing kits. Furthermore, continuous monitoring by humans might harm the medical staffs' mental health~\cite{greenberg2020managing}. 

Mobile applications (apps) are a prevalent solution for active monitoring and individual quarantine for large-scale potential patients in most countries~\cite{peak2020individual}. Although mobile app-based monitoring is effective, it has some limitations. App-based monitoring assumes that the monitoring subjects are familiar with using apps on mobile devices. However, considering that the fatality rate is much higher in older people who are likely to be less familiar with mobile devices~\cite{shim2020transmission}. It is also noticeable that only 37.8\% over-70s in Korea have smartphones according to the survey in 2018. Thus, the assumption of familiarity on app might hinder the efficacy of the app-based solutions. 

To address this issue, we design a call-based artificial intelligence (AI) dialog agent system for monitoring people who contacted injected patients. i.e., CareCall. CareCall consists of three main modules including phone-based automatic speech recognition (ASR)~\cite{chan2016listen, ha2020clovacall}, natural language understanding (NLU), and speech synthesis~\cite{song2019excitnet}, thus calling active monitoring subjects twice per day to check and gather the core symptoms of whether they feel fever and respiratory pains with simple yes or no questions. Because the user experience of CareCall is similar to the human phone-based conversation, our system can track the status of the potential patients more effectively. For enhancing the success ratio, we employ a human-in-the-loop approach in operating our system by monitoring the conversations and adjusting uncertain utterance cases.

We have operated our CareCall system to track active monitoring subjects who contacted the COVID-19 patients in Korea and Japan for three months. In this paper, we introduce a case-study in Seongnam-si, Korea, to show how the system works. In the case-study, CareCall shows 0.9\% as false positive and one case as a false negative. In addition, our human-in-the-loop process remarkably reduces the false positive rate from 1.95\% to 0.72\% when comparing between call cases of the first month and the next two months. Furthermore, we discuss some ideas on how to measure the spreading of the pandemic in the local community based on our system and describe the remaining or newly emerging technical challenges that we are facing, in aspects of the functionalities of call-based AI dialog systems to improve our system. 

\begin{figure*}[t]
\vskip 0.2in
\begin{center}
\centerline{\includegraphics[width=0.7\textwidth]{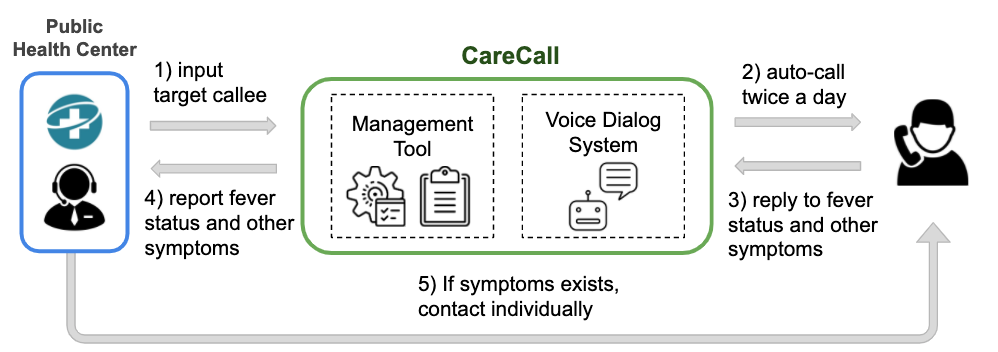}}
\caption{Schematic flow of CareCall system for active monitoring of self-quarantining people.}
\label{fig:fig1}
\end{center}
\vskip -0.3in
\end{figure*}

\section{Related Work}
\subsection{COVID-19}
COVID-19 pandemic has been a critical issue to significantly affect political, social, and economic situations beyond health area around the world. Basically, because COVID-19 is a variant of Corona virus, it is not trivial to distinguish its symptom from respiratory diseases such as a cold. It is frequently reported that the COVID-19 patients suffer from loss of taste and smell functions, which alleviates the difficulty of finding COVID-19 
infected patients ~\cite{gane2020isolated,sungnak2020sars}. However, while 65\% of positive cases on COVID-19 show loss of taste or smell function, 22\% of negative cases also suffer from the same symptom \cite{sungnak2020sars}.

Since infectious disease pandemic leads to enormous loss of various areas, medical scientists have tried to predict epidemic and pandemic for several tens of years. Recent advancement of IT also has enabled to predict diseases without medical equipment such as forecasting flu epidemic from Google search data~\cite{helft2008google, dugas2012google}. However, we cannot assure that keyword search action itself means that they are injected by a disease.

Recently, some apps based on mobile devices are prevalent to track people who contacted COVID-19 patients, who are subjects of social quarantine or active monitoring~\cite{menni2020real}. This work made the app to log health status of individuals, and used 2.6 million people in the UK and the US had been regularly logged their health status. Among these, 18 thousand people reported having had a test for coronavirus, with 7 thousand testings positive. This app-based tracking approach is more effective than conventional search-based methods because the report via app is an explicit action on the disease. However, app-based systems might be ineffective for older people who are not familiar with using mobile devices. Whereas, CareCall uses phone-based communication incorporating speech-based AI dialog models, thus providing more accessibility to older people. 

\subsection{Call-based AI Dialog Agents}
User experiences of call-based tracking systems are different from app-based solutions. Even if the gathered data via app-based systems are likely to be less noisy, call-based system can cover more target subjects who do not use smart devices which are necessary for app operation. In particular, this situation is more serious for older people. Also, a call-based communication is much easier than using apps. However, a call-based system requires more expensive operation costs due to hiring human communicators. Under the pandemic where the monitored people dramatically increase, the cost exponentially increases to track all the subjects. Recent advancement of deep learning has remarkably enhanced the performances of automatic speech recognition~\cite{nassif2019speech}, natural language processing~\cite{devlin2019bert,adiwardana2020towards}, and speech synthesis~\cite{anumanchipalli2019speech}, thus resulting in commercializing call-based AI system~\cite{leviathan2018google}. This call-based AI system can deal with many subjects in practical time as an alternative to an app-based solution.

\section{CareCall}

\subsection{Dialog System}

Our system is a relatively simple task-oriented dialog (TOD) system. Our natural language understanding (NLU) model directly classifies binary slot, and the result corresponds to the system act of the system. There are two explicit slots in our dialog. One is whether the callee has a fever, the other is whether the callee has any respiratory disease. The agent also asks the type of specific symptoms when the callee reported any respiratory disease. However, in this case, the dialog system does not explicitly extract the slot, and just send a dialog log to public servants of the public health center. We use the method similar to the M2M approach to gather data to train our NLU model \cite{shah2018building}.

Why is it sufficient despite our dialog system is simple? There are four reasons. First, the core information we need to extract is simple. The primary goal of our agent is to check the cases where the callee has no symptoms related to COVID-19. Second, in our service situations, if the callee reports any symptoms, the human inspector should read the automatic speech recognition (ASR) dialog log or call again to the callee to ask symptoms again and guide when and where he should go to medical check.  
Third, the system was required to be made as soon as possible. We are requested to make the whole system within three days.
Lastly, it is required to make the number of false positive cases to be minimized. It is critical that the callee reported the symptoms but the dialog system did not catch. 

\subsection{Active Monitoring}

Figure \ref{fig:fig1} describes active monitoring using CareCall.
Unlike other TOD systems, it is extremely important to minimize false negative cases.
In addition to the recognized reports of symptoms, we also use top-1 softmax value and the number of dialog turn as features, and utilize an additional rule-based uncertainty detector with a threshold. 
In other words, not only the system recognizes the callee report symptoms, but also the system think the inference of the system itself is uncertain, CareCall send a log to public servants.
The log is kept for one month before being discarded.

\begin{table}[t!]
\begin{center}
\caption{Turn-level false negative and false positive from rates}
\label{table:table1}
\begin{tabular}{ l r r r r}
\hline
  &\multicolumn{2}{c}{March} & \multicolumn{2}{c}{April-June} \\
  & Count & Ratio & Count & Ratio \\ \hline
 False negative & 0 & 0.00\% & 1 & 0.00\%\\  
 False positive & 88 & 1.95\% & 169 & 0.72\%\\ \hline
 Total turns & 4,508 & 100.00\% & 23,300 & 100.00\%\\ \hline
\end{tabular}
\end{center}
\vskip -0.2in
\end{table}

\subsection{Human-in-the-loop}

It is challenging to make an extremely accurate dialog system using a handful of dataset prepared in the system building phase.
For example, older people have different speech pattern and pronunciation to other groups, with make speech recognition and language recognition challenging.

We can quickly boost the performance of our system using the failed ASR and NLU results by human-in-the-loop. The concept of active learning can be used. Similar to active monitoring, uncertainty for the utterance, like top-1 softmax probability \cite{hendrycks2016baseline} of Bayesian-style uncertainty \cite{siddhant2018deep}, and uncertainty in the dialog-level can be used to find data to be labeled. By labeling a small amount of data active learning suggested, We can easily decrease the error of ASR and NLU module.

\section{CareCall Operation Analysis}

\begin{table}[t!]
\begin{center}
\caption{Conversation examples between CareCall and subjects. U1 refers to cooperative people. U2 refers to people who is hard to interact in the monitoring conversation.}
\label{table:table2}\small
\begin{tabular}{l l | p{6cm}} 
\\
\hline
  \multicolumn{2}{c|}{Speakers}& Utterances \\
  \hline
  \multicolumn{2}{c|}{\multirow{3}{*}{System}} & Hello. I'm calling to check your symptoms regarding to COVID-19 today. Have you got a minute to talk? \\
  \hline
  \multirow{3}{*}{User} & U1 & Hello? \\
  & \multirow{2}{*}{U2} & Yeah, oh, you have already called me earlier today, huh? Yes, go ahead.\\ \hline
  \multicolumn{2}{c|}{System} & Hello again. Is this a good time to talk? \\ \hline
  \multirow{2}{*}{User} & U1 & Yes. \\
   & U2 & Yes. \\ \hline
  \multicolumn{2}{c|}{System} & Do you have a fever now? \\ \hline
  \multirow{3}{*}{User} & U1 & No. \\
  & \multirow{2}{*}{U2} & Yeah. Nothing like that. I'll let you know if there's anything like that. Oh. Too stressful. \\ \hline
  \multicolumn{2}{c|}{\multirow{3}{*}{System}}  & Okay. Do you have a cough or symptoms like shortness of breath now? Please answer yes or no. \\ \hline
  \multirow{2}{*}{User} & U1 & No. I don't \\
  & U2 & I am totally fine. Please do not worry. \\ \hline
  \multicolumn{2}{c|}{\multirow{4}{*}{System}} & Okay. Got it. When you want to go outside, be sure to wear your mask. If you think you have any suspect symptoms, please contact the public health center. Thank you. \\
  \hline
\end{tabular}
\end{center}
\vskip -0.2in
\end{table}

We have released CareCall to monitor the citizens whose symptoms related to COVID-19 are needed to be checked. CareCall has been operating in Seongnam-si, Korea, since March 2020. Our system helps reduce the burden of monitoring work on nearly one-thirty of the total monitoring needs. Figure 2 shows the actual confirmed cases of COVID-19 in Seongnam-si. Since March 9th, 142 confirmed cases have been reported and one patient has died among them. We analyzed data from CareCall to improve the performance of our dialog system.

All data including quantitative data and call logs from CareCall were analyzed to understand the interaction between the subjects and our system. Call hang-up rate by the subjects before completing the conversation is 14.6\%, and connection failure rate is 7.3\%. Those two rates are relatively low because monitored subjects are responsible for receiving the monitoring call. 

We also investigated turn-level errors in the entire monitoring cases. Our target data were logs from a total of 13,904 calls. We analyzed each turn-level false negative (FN) and false positive (FP) cases (see Table 1). The false negative case means that monitored subjects report the COVID-19 symptoms but CareCall does not confirm it, and this case could be critical but only one case occurred for three months. This case was escalated to human monitor of COVID-19 in Seongnam-si. On the contrary, the false positive means CareCall detects the symptoms of monitored subjects although they have no symptom; the false positive rate (FPR) is 0.92\%, which is very low as well. Based on the data analysis, we could improve the performance of CareCall (see Table 1). Remarkable improvement of performance from April results from NLU and speech synthesis model update and data refinement by our human-in-the-loop process. 

\begin{figure}[t]
\vskip 0.2in
\begin{center}
\centerline{\includegraphics[width=\columnwidth]{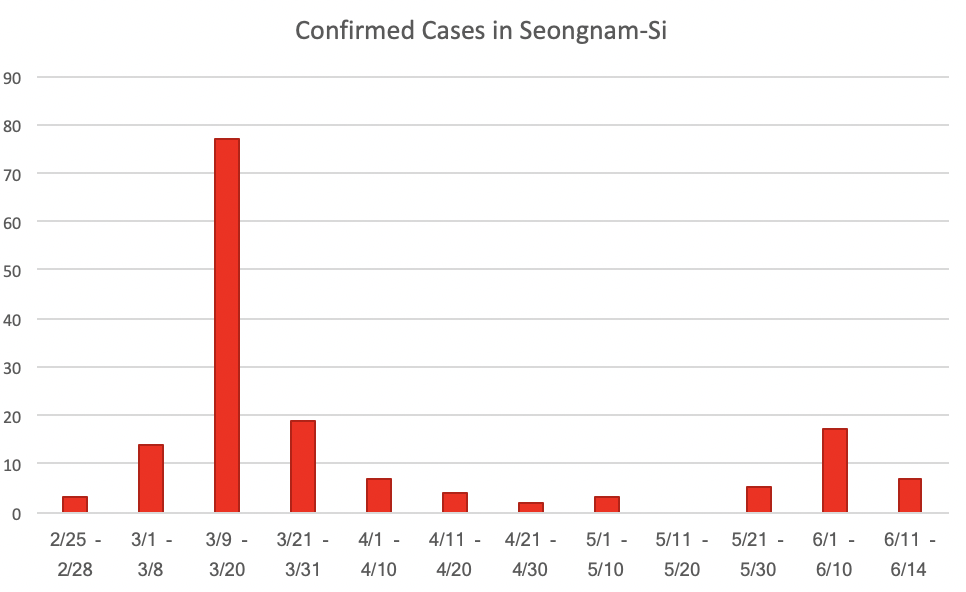}}
\caption{Confirmed cases in Seongnam-si until Jun 15th. March 9th was the critical period for monitoring subjects in Seongnam-si. CareCall released at the right timing to reduce the burden of monitoring case.} 
\label{fig:confirmed}
\end{center}
\vskip -0.4in
\end{figure}

CareCall asks polar questions to monitored subjects, and they need to answer simply `yes' or `no' to the questions. Most of the monitored subjects could easily interact with the voice agent of CareCall. However, since older people tended to respond more freely, it was difficult for the dialog system to classify the utterances of older people. This is a challenging technology issue we need to tackle. Firstly, a voice-based dialog system is required to be able to understand unexpected type of user utterances. Therefore NLU module could be crucial in this voice-based interface. Furthermore, handling utterances of older people could be challenging because they easily expressed their emotion against the system although the voice agent of CareCall was not a real human \cite{leviathan2018google}. However, older people is an important group in COVID-19 monitoring.
In the statistics in Korea until early June,
over-60s accounts for more than 23\% of confirmed cases and 92\% of death cases. Therefore, it is crucial to monitor older people and it should not be excluded from the investigation as an exceptional case.

\section{Discussion}
\subsection{Pandemic Spread Prediction Models}

We discuss an idea on how to measure spreading of the pandemic based on our system in terms of simple probabilistic modeling.
It is important to prevent community spread by tracking, and some countries are successfully preventing a serious situation. In the situation where local community spread is suspected, tracking and estimating the spread status makes some advantageous for deciding to extend self-quarantine policy or execute other helpful actions. 
Specifically, a statistically significant infection rate and the number of estimated infected patients would be helpful.
To evaluate the individual infection, a previous work uses linear regression with the person's symptoms as a feature \cite{menni2020real}. However, the local community's infection rate is also a critical factor in evaluating individual infection.
To this end, our idea is to ask the symptoms to not only self-container but also other random people in the community to estimate how severe the spread in the community is.

\begin{figure}[t]
\vskip 0.2in
\begin{center}
\centerline{\includegraphics[width=0.8\columnwidth]{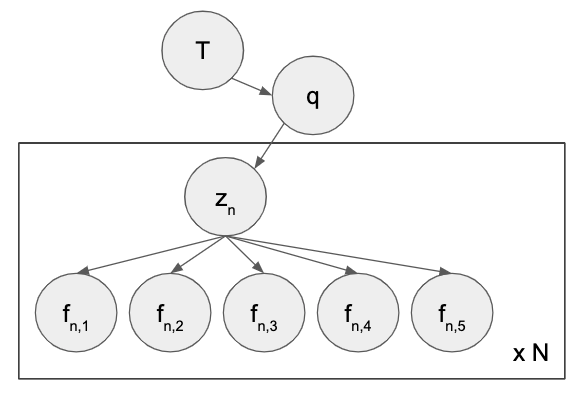}}
\caption{An example of PGM for modeling a infection rate $q$ and an individual infection $z_n$}
\label{fig:pgm}
\end{center}
\vskip -0.4in
\end{figure}

To aggregate the reported symptoms from callees, a Bayesian approach can be used for modeling spread degrees in a local community. 
The Bayesian approach utilizes the information from statistics of previous COVID-19 spread as the prior to estimate the posterior probability of community infection rate by using the symptoms of people investigated.

Figure \ref{fig:pgm} presents a simple example of Bayesian modeling. $T$ is defined as binary value where 1 denotes infection case exists, whereas 0 denotes it does not exist. $0 \leq q \leq 1$ is a continuous value, which denote infection probability. $q$ and $T$ are modeled as separated random variables because we want to model $p(q=0)$ like delta-function. Otherwise, $\frac{p(q=0)}{p(q>0)} = 0$. Prior of $T$ and $q$ (i.e., $p(T)$ and $p(q|T)$) can be characterized by the statistics of previous COVID-19 spread.

Individual infection $z_n$ is a binary value, which denotes whether individual $n$ is infected or not. A confirmed case can be considered as $z_n = 1$. $f_{n,v}$ is the $n$-th individual's feature $v$. For example, a feature can be loss of smell and taste. $F=\{F_1, \cdots, F_n, \cdots, F_N\}$, and $F_n = \{f_{n,1}, \cdots, f_{n,v}, \cdots, f_{n,V}\}$.
We can also define $p(z_n|q) = q$ and $p(F_n|z_n) = \prod_i p(f_{n,i}|z_n)$. In this formulation, calculating the posterior of infection rate $p(q|F)$ or an individual infection $p(z_n|F)$ would be one of primary interests.

\subsection {Extension to General Healthcare Recommender Systems}
Extending our CareCall system to more general healthcare domains is natural because the system flow is not specific to COVID-19 only. First, our system can be extended to making a call-based examination for cold or flu without heavy additional efforts. By the examination results, our system can recommend whether users should go to see a doctor or which department they need to go to. Furthermore, we can also extend this system to cover more diseases by customizing the graph definition of dialog state and action corresponding to intents, slots, human-to-human dialog utterances, and disease-question-answering data. In particular, CareCall might be effective for managing chronic diseases such as diabetes, hypertension, and hyperlipidemia which many elderly people not familiar with mobile applications suffer from. Our system can help the elderly patients by monitoring the status of patients and recommending the actions corresponding to the patient state.

\section{Conclusion}
We introduced CareCall, a call-based AI dialog agent system for monitoring people who contacted injected patients, and showed how this system is built and works robustly.
We also discussed the idea of applying our system to measure the spreading of the pandemic in the local community with simple probabilistic modeling.
We hope this kind of investigation actively plays a role in tracking and preventing the COVID-19 spread.

\section*{Acknowledgements}
The authors thank all members of Clova CIC for supporting this work. Also, the authors appreciate all the medical staffs around the world for their devoted efforts to prevent COVID-19.

\bibliography{example_paper}
\bibliographystyle{icml2020}

\end{document}